%Paper: hep-th/9404074
%From: Simon Brian Davis <S.B.Davis@damtp.cambridge.ac.uk>
%Date: Wed, 13 Apr 94 18:34:23 BST
%Date (revised): Mon, 25 Apr 94 13:30 BST
%Date (revised): Fri, 22 Apr 94 18:06:29 BST

\magnification=1200
\parskip=10pt plus 5pt
\parindent=12pt
\baselineskip=18pt
\input mssymb
\pageno=0
\footline={\ifnum \pageno <1 \else \hss \folio \hss \fi }
\line{\hfil {DAMTP-R/94/16}}
\vskip .65in
\centerline{\bf CONFIGURATIONS OF HANDLES AND THE CLASSIFICATION OF}
\vskip 2pt
\centerline{\bf DIVERGENCES IN THE STRING PARTITION FUNCTION}
\vskip .65in
\centerline{Simon Davis}
\vskip .4in
\centerline{Research Foundation of Southern California}
\vskip 1pt
\centerline{5580 La Jolla Boulevard, La Jolla, CA 92037}
\vskip 2pt
\centerline{and}
\vskip 2pt
\centerline{Department of Applied Mathematics and Theoretical Physics}
\vskip 1pt
\centerline{University of Cambridge}
\vskip 1pt
\centerline{Silver Street, Cambridge CB3 9EW \footnote{*}{Present address}}
\vskip .5in
{\bf Abstract}.  The divergences that arise in the regularized partition
function for closed bosonic string theory in flat space lead to three types of
perturbation series expansions, distinguished by their genus dependence.
This classification of infinities can be traced to geometrical characteristics
of the string worldsheet.  Some categories of divergences may be eliminated in
string theories formulated on compact curved manifolds.
\vfill
\eject

$\underline{1.~Introduction}$

The resolution of fundamental problems in string theory such as the
determination of the stability of phenomenologically realistic vacua and the
existence of black hole solutions, might require a complete formulation of
the theory including non-perturbative effects as well as S-matrix
calculations [1][2].  A study of the perturbation series, however, may be
sufficient to establish stability of the vacuum.  For closed bosonic strings
moving in 26-dimensional flat space-time, the infrared divergences in the
partition function can be eliminated through integration over a subset of
moduli space that does not include the degeneration locus.  Although an
estimate has already been made for the genus-dependent growth of integrals
with this cut-off over Teichmuller space, it will be altered by the restriction
from Teichmuller space to moduli space [3].  A precise definition of the
integration region is necessary to establish rigorously that the flat
background
is not stable against perturbations.

Through an analysis of the conditions defining the fundamental region of the
modular and mapping class groups, a lower bound for the regularized partition
function has been found [3][5].  While the action of the symplectic modular
group Sp(2g;${\Bbb Z}$) is defined initially on a ${1\over
2}$g(g+1)-dimensional space
of positive-definite symmetric matrices, the fundamental domain can be
projected
onto a (3g-3)-dimensional subspace of period matrices $\tau$ for Riemann
surfaces of genus g satisfying the inequalities
$$\eqalign{(i)&~~-{1\over 2}~\le~Re~\tau_{mn}~\le~{1\over 2}
\cr
(ii)&~~\vert det(C \tau+D) \vert~\ge~1~~~~\left(\matrix{A&B&
                                                   \cr
                                                  C&D&
                                                  \cr}\right)\in Sp(2g;{\Bbb
Z})
\cr
(iii)&~~Im~\tau[g_r]~=~g_r^T(Im~\tau)g_r~\ge~Im~\tau[e_r]~where~e_r=(0,...,0,1,0,...,0)
\cr
&~and~g_r=(g_{r1},...,g_{rr},...,g_{rg})~where~g_{rr},...,~g_{rg}~are~
relatively
\cr
&~prime;~(Im~\tau)_{1r}~\ge~0
\cr}$$
The bound for the partition function has been obtained, using another set of
variables, the multipliers $\{K_n\}$ and the fixed points
$\{\xi_{1n}, \xi_{2n}\}$, characterizing the generators $\{T_n,n=1,...g\}$ of
the Schottky group
uniformizing the Riemann surface.  In particular, limits have been placed
on the multipliers and fixed points using the relation between the elements of
the period matrix and these variables [4],
$$\tau_{mn}~=~{1\over {2\pi
i}}~\left[~ln~K_m~\delta_{mn}~+~\sum_\alpha~^{(m,n)}
ln~{{\xi_{1m}~-~V_\alpha \xi_{1n}}\over {\xi_{1m}~-~V_\alpha \xi_{2n}}}
{{\xi_{2m}-V_\alpha \xi_{2n}}\over {\xi_{2m}-V_\alpha \xi_{1n}}}~\right]$$
with $V_\alpha$ being a product of $T_{n_i}$ and $\sum_\alpha~^{(m,n)}$
excluding all $V_\alpha$ containing $T_m^{\pm 1}$ at the left and $T_n^{\pm 1}$
at the right.

The vacuum amplitudes at finite order may be factored out of the higher-point
correlation functions as the closed Riemann surfaces are equivalent to the
bubble diagrams of field theory.  Because of the factorization of the closed
string vacuum amplitude, the physical significance of the cut-off integral over
moduli space is revealed in the effect of the summability of the series
on the stability of the classical background.  Moreover, it allows one to
study the limit as the number of handles tends to infinity, where the surfaces,
depending on the classification type, may represent the creation of a
point-particle or string state from the vacuum.  The four-point function on
an infinite-genus surface of type $O_G$ has been evaluated in previous work
[6].

Surfaces containing an infinite number of accumulating handles affect the
Borel summability of the perturbation series.  It will be demonstrated that
there are three categories of surfaces associated with distinct arrangements
of the isometric circles $\{I_{T_n}\}$ of the generators in the complex plane.
The growth of both the lower and upper bounds is related to the geometrical
characteristics of the configuration of isometric circles or equivalently the
handles on the surfaces.

For strings propagating in flat space-time, the growth of the bound for the
partition function is obtained after an analysis of each of the configurations
of isometric circles which lie in the fundamental region of the symplectic
modular group [7].  The three categories considered are determined by the
the dependence of the multipliers and fixed-point distances on the genus, which
are
related because
 ${{\vert K_n\vert}\over {\vert 1-K_n\vert^2}}={{\vert\gamma_n\vert^{-2}}\over
{\vert\xi_{1n}-\xi_{2n}\vert^2}}$
and
 $min_n \vert\gamma_n\vert^{-2}\sim {1\over g}$ as a consequence of requiring
genus-independence of the cut-off [9].  If
$\vert \xi_{1n}-\xi_{2n}\vert\sim
{1\over {g^q}}$, the values q=0, $q={1\over 2}$,  and $0<q<{1\over 2}$
define the three types of configurations.   It is demonstrated in this paper
that the contributions of the correponding surfaces to the string partition
function will increase as
$k^g {{g^g}\over {(ln~g)^{13g}}}$, $k^{\prime g}({g\over J})!$
and $k^{\prime\prime g}\left({{g^g}\over {(ln~g)^{13g}}}\right)$ respectively.
These estimates follow from an evaluation of upper and lower
bounds for the integrals with limits for the Schottky group parameters
corresponding  to the different values of q.  This establishes rigorously that
an approximately factorial growth
will occur in the regularized string partition function,
a property found in previous work [2][3] for the string integral over subset of
Teichmuller space without restriction to the fundamental domain of the mapping
class group.

 Since each of the terms in the perturbation series is positive,
being the volume of a region in a (3g-3)-dimensional parameter space, an
exactly factorial increase,
$k^g g!$ would indicate that the series is not Borel summable.  The growth
obtained here for specific values of q is not exactly factorial, so that its
effect on the Borel summability of the series would require further analysis.
For the three categories of isometric circles studied here, the modified
series, defined by dividing the gth term by g!, would be convergent on the
entire
positive real axis in the coupling constant plane, allowing for the use of the
Borel integral transform.  However, summation of the contributions of the
three kinds of configurations to the partition function only comprises part of
the regularized moduli space integral.  Specifically, the third category can be
subdivided into a further $O(ln~g)$ sub-categories, and by considering
configurations of isometric circles with multipliers $\{K_n\}$  having
different
genus dependence for different n, an additional combinatorial factor of the
form
${{(g+k~ln~g-1)!}\over {g!(k~ln~g-1)!}}$
is introduced.  While this improves
the rapidly increasing growth of the bounds for the regularized integral, it
is still not large enough to alter the Borel summability properties of the
series.  Additional contributions to the string partition function
arise from special configurations within the three categories that lead to a
change in the genus-dependence of certain factors in the moduli space integral.
 By applying the
techniques used here to these configurations, the magnitude of the
contributions may be estimated, and it appears that they will not significantly
change the
bounds.  Finally, one might investigate other categories of
isometric circles consistent with a genus-independent cut-off on the length
of closed geodesics on the surface.  A discussion of a separate category of
\vfill\eject
\noindent
isometric circles is given in Section 3, and a complete study of the other
configurations may be required to find a precise estimate of the regularized
string integral in the Schottky group parametrization.

The results obtained in this paper indicate that it may be useful to re-examine
the application of graph theory [3], used to obtain a factorial growth with
respect to the genus, in the large-order limit.  In particular, it is suggested
that a there is a potential difference between the
counting of diagrams corresponding to string  and point-particle trajectories
and this would be determined by the physical intepretation of string theory in
the large-genus limit.

The change of metric in a curved manifold leads to a different bosonic string
path integral.  The divergences in the partition function for strings
propagating in a curved background are considered, and an outline of the
analysis of the allowed configurations of isometric circles in a compact target
space and their effect on Borel summability is presented.
\vskip 20pt
$\underline{2.~Divergences~in~Flat~Space~Closed~Bosonic~String~Perturbation~
Theory}$

By integrating over metrics on the string worldsheet and target space
coordinates in ${\Bbb R}^{26}$, the partition function for the closed string
theory,
obtained by summing over the genus,
\vskip 10pt
$$Z~=~\sum_{g=0}^{\infty}~\kappa^g~\int{{D[h_{\alpha\beta}]D[X_\mu]}
\over {Vol(Diff~\Sigma_g) Vol(Conf~\Sigma_g)}}e^{-{1\over 2}\int~d^2 \xi
\sqrt{h} h^{\alpha\beta} \partial_\alpha X^\mu \partial_\beta X_\mu}
\eqno(1)$$
\vskip 10pt
\noindent
can be re-expressed in terms of a measure involving Schottky group variables
$K_n$, $\xi_{1n}$ and $\xi_{2n}$, which are the multipliers and fixed points
defining
the generating transformations of the uniformizing group through
${{T_n z-\xi_{1n}}\over {T_n z-\xi_{2n}}}=K_n {{z-\xi_{1n}}\over
{z-\xi_{2n}}}$.
The partition function [8] is given
\vfill\eject
\noindent
by
$$\eqalign{Z~=~\sum_{g=0}^\infty~c_g
\kappa^g~\int_{F_g}~\prod_{m=1}^g~&{{d^2K_m}\over
{\vert K_m\vert^4}}\vert 1-K_m\vert^4 {1\over {Vol(SL(2,{\Bbb
C}))}}~\prod_{n=1}^g
{}~{{d^2 \xi_{1n} d^2 \xi_{2n}}\over {\vert \xi_{1n}-\xi_{2n}\vert^4}}
\cr
\cdot& [det(Im \tau)]^{-13}
\prod_\alpha~^{\prime}\prod_{p=1}^\infty
\vert 1-K_\alpha^p\vert^{-48} \prod_\alpha~^{\prime} \vert 1-K_\alpha\vert^{-4}
\cr}
\eqno(2)$$
where $\tau$ is the period matrix, $K_\alpha$ is the multiplier of the element
$V_\alpha$, $\prod_\alpha^{\prime}$ represents the product over conjugacy
classes of primitive elements, and $F_g$ is the fundamental region of the
modular group in the parameter space.  In earlier work [5], a lower bound was
set for each of the terms in the integrand in equation (2), when the Riemann
surface
$\Sigma_g$ is constrained by the requirement that the minimal length of closed
geodesics on the manifold be bounded below.

The action of the generator $T_n$, $T_n z={{\alpha_n z+\beta_n}\over {\gamma_n
z
+\delta_n}}$, on the complex plane involves the mapping of the exterior of
$I_{T_n}=\{z\in {\Bbb C}\vert \vert\gamma_n z+\delta_n\vert=1\}$ to the the
interior
of $I_{T_n^{-1}}=\{z\in {\Bbb C}\vert \vert\gamma_n z-\alpha_n\vert=1\}$ and
the interior of $I_{T_n}$ to the exterior of $I_{T_n^{-1}}$.  There are three
types
of configurations of isometric circles determined by the behaviour of the
multipliers and fixed points consistent with the cut-off
$\vert\gamma_n\vert^{-2} \sim {1\over g}$.  Other configurations of isometric
circles consistent with this decrease in the radii of circles on a sphere of
projection are considered in the following section.  The three categories of
isometric circles studied in this section are characterized by the dependence
of the limits on the genus.
$$\eqalign{(i)&~~{{c_1}\over g} \le \vert K_n\vert \le {{c_2}\over
g}~~~~~~~~~~~~~
\delta_0 \le \vert\xi_{1n}-\xi_{2n}\vert \le \delta_0^{\prime}
\cr
(ii)&~~\epsilon_0\le \vert K_n\vert\le \epsilon_0^\prime~~~~~~~~~~~~~
{{\delta_0}\over {\sqrt{g}}}+d(\xi_{1n},
I_{T_n^{-1}})\le\vert\xi_{1n}-\xi_{2n}\vert\le {{\delta_0^{\prime}}\over
{\sqrt{g}}}
\cr
(iii)&~~{{\epsilon_0}\over {g^{1-2q}}}\le \vert K_n\vert\le
{{\epsilon_0^{\prime}}\over {g^{1-2q}}}~~~
{{\delta_0}\over {g^q}}+d(\xi_{1n}, I_{T_n^{-1}})\le
\vert\xi_{1n}-\xi_{2n}\vert
\le {{\delta_0^{\prime}}\over {g^q}}~~~~0<q<{1\over 2}
\cr}$$
Bounds for the integrals over subsets of moduli space corresponding to the
first two types  of configurations have been found [5].  In the following
analysis, these bounds are refined and an estimate of the contribution of the
third
type of configuration is given.

By setting an upper bound on $\sum_\alpha~^\prime \vert K_\alpha\vert=
\sum_\alpha~^\prime {{\vert\gamma_\alpha\vert^{-2}}\over {\left\vert
\xi_{1\alpha}
+{{\delta_\alpha}\over {\gamma_\alpha}}\right\vert^2}}$, the next two
inequalities have
been obtained for isometric circles defined by the limits in category (i).

$$\eqalign{\prod_\alpha~^\prime\prod_{p=1}^\infty
\vert 1-K_\alpha^p\vert^{-48}
\prod_\alpha~^\prime\vert 1-K_\alpha\vert^{-4}~>~
& exp\left({{-192c^4}\over {c^2-2c_2}}{{d_{max}^2(\{I_{T_n},I_{T_n^{-1}}\})}
\over {d_{min}^2(\{I_{T_m}\})}}\right)
\cr
&\left({1\over 2}\right)^{\left(1+{{192}\over {c_2}}
{{d_{max}^2(\{I_{T_n}, I_{T_n^{-1}}\})}\over {d_{min}^2(\{I_{T_m}\})}}
((c_2-c^2-c+{1\over 2})C_0-C_1)\right)}
\cr}
\eqno(3)$$

$$\eqalign{C_0~=&~{{c^4}\over {c^2-2c_2}}
\cr
C_1~=&~-c_2c^3{{(c-4c_2-4c_2c)}\over {(c^2-2c_2)^2}}
\cr
C_2~=&~{{c^2(1+c)c_2^2}\over {c^2-2c_2}}~+~{{2c^3(1+c)c_2^2-5c^4(1+c)^2c_2^2}
\over {(c^2-2c_2)^2}}~+~{{c^4c_2^2(1-2c(1+c))^2}\over {(c^2-2c_2)^3}}
\cr
{}.
\cr
{}.
\cr
{}.
\cr}
\eqno(4)$$
with c being a lower bound for $\left\vert{{\xi_{2n}-{{\alpha_\beta}\over
{\gamma_\beta}}}\over {\xi_{2n}-\xi_{1n}}}\right\vert$ for bounded g, when the
distance between the isometric circles is O(1), and a
genus-independent constant for large g, $I_{V_\beta^{-1}}
\not\subset D_{T_n}=\{z\vert\vert\gamma_n z+\delta_n\vert \le 1\}$,
$d_{min}(\{I_{T_m}\})$ equal to the minimum distance between the isometric
circles and $d_{max}(\{I_{T_n}, I_{T_n^{-1}}\})$ equal to the maximum distance
between $I_{T_n}$ and $I_{T_n^{-1}}$, for all n;
$$[det (Im~\tau)]^{-13}~>~(2 \pi)^{13g}\left[ln~g+{{8c}\over {c^2-2c_2}}-ln~c_1
\right]^{-13 g}
\eqno(5)$$
Configurations of isometric circles in each of the three categories satisfy a
set of conditions defining a fundamental region of the symplectic modular group
acting on a (3g-3)-dimensional subset of the Siegel upper-half space if
appropriate  restrictions are placed on the multipliers $K_n$ and fixed points
$\xi_{1n}$ and $\xi_{2n}$.  The fundamental set of inequalities result from
$Im~\tau[g_r]~\ge~Im~\tau[e_r]$ when $g_r=e_s,~s\ge r$.  It has been shown that
these inequalities imply the conditions for a general column vector $g_r$ at
sufficiently large genus [5].
     For isometric circles of the first type, the inequalities $(Im \tau)_{ss}
\ge (Im \tau)_{rr},~s\ge r$, lead to a reduction
of the integration range of the multipliers.  Denoting
$$\eqalign{\rho^{-1}~=&~max_n~\prod_\alpha~^{(n,n)}\left\vert
1-{{\gamma_\alpha^{-2}
(\xi_{1n}-\xi_{2n})^2}\over {\left(\xi_{1n}+{{\delta_\alpha}\over
{\gamma_\alpha}}\right)\left(\xi_{2n}+{{\delta_\alpha}\over
{\gamma_\alpha}}\right)(\xi_{1n}-V_\alpha\xi_{1n})(\xi_{2n}-V_\alpha\xi_{2n})}}
\right\vert
\cr
\sigma~=&~min_n~\prod_\alpha~^{(n,n)}\left\vert 1+{{\gamma_\alpha^{-2}(\xi_{1n}
-\xi_{2n})^2}\over {\left(\xi_{1n}+{{\delta_\alpha}\over {\gamma_\alpha}}
\right)\left(\xi_{2n}+{{\delta_\alpha}\over {\gamma_\alpha}}\right)
(\xi_{1n}-V_\alpha \xi_{2n})(\xi_{2n}-V_\alpha \xi_{1n})}}\right\vert
\cr}
\eqno(6)$$
the range of multipliers $K_1,...,~K_g$ can be selected to be
$[\rho^{-2}{{c_1}\over g},~\sigma^2 {{c_2}\over g}],~[\rho^{-2}{{c_1}\over g},
\sigma^2 \vert K_1\vert]$
,...,$[\rho^{-2}{{c_1}\over g}, \sigma^2 \vert K_{g-2}
\vert],~[{{c_1}\over g},\sigma^2 \vert K_{g-1}\vert]$, so that the integral
over the multipliers will be reduced to
$$\eqalign{(2 \pi)^g \int_{\rho^{-2} {{c_1}\over g} }^{{c_2}\over g}&~
\int_{\rho^{-2}{{c_1}\over g}}^{\sigma^2 \vert K_1\vert}~...~
\int_{\rho^{-2} {{c_1}\over g}}^{\sigma^2 \vert K_{g-2}\vert}~
\int_{{c_1}\over g}^{\sigma^2 \vert K_{g-1}\vert}~{{d\vert K_g\vert}\over
{\vert
K_g\vert^3}}{{d\vert K_{g-1}\vert}\over {\vert K_{g-1}\vert^3}}...
{{d\vert K_2\vert}\over {\vert K_2\vert^3}}{{d\vert K_1\vert}\over {\vert
K_1\vert^3}}
\cr
{}~=&~(2 \pi)^g~\sum_{j=0}^g~\kappa_{g-j}(\sigma,\rho)\left[1-{{c_1^2}
\over {\rho^4\sigma^{4(g-j)} c_2^2}}\right]^{g-j} {{\rho^{4(g-j)}g^{2(g-j)}}
\over {2^{g-j} c_1^{2(g-j)}}}
\cr}
\eqno(7)$$
$$\eqalign{\kappa_g(\sigma, \rho)~=&~{{\sigma^{2g(2g-1)}}\over {g!}}
\cr
\kappa_{g-1}~=&~0
\cr
\kappa_{g-2}(\sigma, \rho)~=&~-{{\sigma^4 \rho^8 g^4}\over {c_1^4}}
\left[1-{1\over {\sigma^4}}\right]^2 {{\sigma^{2(g-2)(g-3)}}\over
{(g-2)!}}
\cr
\kappa_{g-3}(\sigma, \rho)~=&~-{{\sigma^{12}\rho^{12} g^6}\over {3! 8c_1^6}}
\left[1-{1\over {\sigma^8}}\right]^3 {{\sigma^{2(g-3)(g-4)}}\over {(g-3)!}}
\cr
.&
\cr
.&
\cr
.&
\cr}
\eqno(8)$$

For a configuration of 2g isometric circles involving a minimal spacing
$d_{min}(\{I_{T_m}\})=O({1\over {\sqrt{g}}})$ and contained in a circular
region of radius $O(1)$, the ratio $\left\vert {{\xi_{1n}-\xi_{2n}}\over
{\xi_{1n}+{{\delta_{T_{n_l}}}\over
{\gamma_{T_{n_l}}}}}}\right\vert=O(\sqrt{g})$
for isometric circles $I_{T_{n_l}}$ near $\xi_{1n}$, and
$\left\vert {{\xi_{1n}-\xi_{2n}}\over {\xi_{1n}+{{\delta_\alpha}\over
{\gamma_\alpha}}}}\right\vert=O(1)$ for circles $I_{V_\alpha}$ near
the boundary of the region.  It follows that the largest remainder terms
in $\rho^{-1}$ and $\sigma$ will be of order $O({1\over {\sqrt{g}}})$, since
$\vert\gamma_{n_l}\vert^{-2}=O({1\over g})$.  However, in a symmetrical
configuration of isometric circles about $\xi_{1n}$, for every circle
$I_{T_{n_{l_1}}}$ near $\xi_{1n}$, there may be another circle
$I_{T_{n_{l_2}}}$
in the same neighbourhood lying on the opposite side of $\xi_{1n}$.  If the
arguments of $\gamma_{n_{l_1}}$ and $\gamma_{n_{l_2}}$ are equal, then
the product of the two corresponding terms in equation (6) would have the
form $1-O({1\over g})$.  Since the products in (6) are defined over an infinite
number of isometric circles, the phases of $\gamma_\alpha^{-2}$ will be
randomly distributed in the interval $[-\pi,\pi]$ and many of the terms of
order $O({1\over {\sqrt{g}}})$ may cancel, leaving a remainder term of order
$O({1\over g})$.
When $\sigma=1-{{\sigma_1}\over g}+{{\sigma_2}\over {g^2}}+...$, it can
be demonstrated, that in the limit as $g\to\infty$, the sum in equation (7) is
$$\eqalign{e^{-2\sigma_1 g}{{\rho^{4g} g^{2g}}\over {2^g c_1^{2g} g!}}
\biggl[e^{2\sigma_1}&\left[1-{{c_1^2\sigma^4 e^{4\sigma_1}}\over {\rho^4
c_2^2}}
\right]^g~-~8e^{10 \sigma_1} \sigma^{16} \sigma_1^2 \left[1-{{c_1^2\sigma^{12}
e^{4\sigma_1}}\over {\rho^4 c_2^2}}\right]^{g-2}
\cr
{}~+&~{{256}\over 3}e^{14 \sigma_1} \sigma^{36} \sigma_1^3\left[1-{{c_1
\sigma^{16}e^{4\sigma_1}}\over {\rho^4 c_2^2}}\right]^{g-3}~+~...~\biggr]
\cr}
\eqno(9)$$
As the factor associated with the jth term has a dependence on j given by
\hfil\break
$(4e)^j e^{2\sigma_1(2j+1)}\sigma^{4j^2}$, the series can be truncated
essentially at j=O(1), and a lower bound of the form $k_1k_2^g$ can be used for
the sum in equation (8).  As the integral over the multipliers is greater than
${{k_1}\over {\sqrt{2\pi g}}}\left({{\pi k_2 e^{1-2\sigma_1} \rho^4}
\over {2c_1^2}}\right)^g g^g$, it can be combined with the inequality for
$[det (Im~\tau)]^{-13}$ and the fixed point integral to give
$$\eqalign{{{k_1 e^{-4c_2}}\over 4}&{{\vert\xi_{11}^0-\xi_{1g}^0\vert^2 \vert
\xi_{21}-
\xi_{1g}^0\vert^2}\over {\vert \xi_{11}^0-\xi_{21}^0\vert^2}}
exp\left({{-192c^4}\over {c^2-2c_2}}{{d_{max}^2(\{I_{T_n}, I_{T_n^{-1}}\})}
\over {d_{min}^2(\{I_{T_m}\})}}\right)
\cr
&\left({1\over 2}\right)^{1+{{192}\over {c_2}} {{d_{max}^2(\{I_{T_n},
I_{T_n^{-1}}
\})}\over {d_{min}^2 (\{I_{T_m}\})}}[c_2C_0+C_1-C_0(c^2+c-{1\over 2})]}
\cr
&g^{-{1\over 2}} (2 \pi)^{16g-3} \left({1\over {\delta_0^2}}-{1\over
{\delta_0^{\prime 2}}}\right)^{g-1} (\delta_2^{\prime 2}-\delta_2^2)^{g-2}
\left({{k_2 e^{1-2\sigma_1} \rho^4}\over {2c_1^2}}\right)^g
\cr
&{{g^g}\over {\left[ln~g+{{8c}\over {c^2-2c_2}}-ln c_1\right]^{13 g}}}
\cr}
\eqno(10)$$
which is a refinement of the bound obtained in earlier work [9].

It has already been noted that the largest remainder terms in equation (6)
would be of order $O({1\over {\sqrt{g}}})$ for configurations of isometric
circles of type (i).  As these terms may not necessarily cancel at this order,
an improved definition of the allowed ranges of values of the variables
$\vert K_n\vert$ is required.
In particular, upon using the formula for the period matrix elements, the
inequalities $(Im~\tau)_{ss}\ge (Im~\tau)_{rr},~s\ge r$ only
lead to constraints of the form ${{\vert K_1\vert}\over {\rho_1}} \ge
{{\vert K_2\vert}\over {\rho_2}}\ge ...\ge {{\vert K_g\vert}\over {\rho_g}}$
where
$\rho_n=\prod_\alpha~^{(n,n)}\left\vert{{\xi_{1n}-V_\alpha\xi_{2n}}\over
{\xi_{1n}-V_\alpha \xi_{1n}}}{{\xi_{2n}-V_\alpha\xi_{1n}}\over
{\xi_{2n}-V_\alpha \xi_{2n}}}\right\vert$.  Whereas integration of the absolute
values $\vert K_n\vert$ over the range
$[{{c_1}\over g},~{{c_2}\over g}]$ gives ${1\over {2^g}}\left({1\over {c_1^2}}
-{1\over {c_2^2}}\right)^g g^{2g}$, integration over the restricted range
is bounded by ${1\over {2^g \rho_1^2...\rho_g^2 g!}} \left({1\over
{\rho^{-2}c_1^2}}-{1\over {\rho_1^2 c_2^2}}\right)^g g^{2g}$.  The modification
associated with the dependence of the remainder terms in the product $\rho_n$
being $O({1\over {\sqrt{g}}})$ does not affect the dominant behaviour of the
bound, which is equivalent to that obtained in equation (10).

In the second type of configuration, all 2g circles can be contained in a disk
of finite radius of order $O({1\over {\sqrt{g}}})$.  The dominant contribution
to the partition function is the fixed point integral
$$\eqalign{\int{1\over {Vol(SL(2,C))}}~\prod_{n=1}^g~&{{d^2\xi_{1n} d^2
\xi_{2n}}
\over {\vert\xi_{1n}-\xi_{2n}\vert^4}}
\cr
&~\ge~{{\vert\xi_{11}^0-\xi_{1g}^0\vert^2\vert\xi_{1g}^0-\xi_{21}^0\vert^2}\over
{\vert \xi_{11}^0-\xi_{21}^0\vert^2}}
\pi^{2g-3} \delta_2^{\prime 2g-4} \left[1-{{(\delta_0+\delta_0^{\prime})^2}
\over {4 \delta_2^{\prime 2}}}\right]^{g-2}
\cr
&~~~~\left[{1\over {\delta_0^2}}\left[1+{{\epsilon_0^{1\over 2}}\over
{1+\epsilon_0}}
\left[1-\sum_{n^{\prime}=1}^\infty {{(2n^\prime -1)!!}\over {(n^\prime +1)!}}
{1\over {2^{n^\prime}}}\right]\right]^{-2} -{1\over {\delta_0^{\prime 2}}}
\right]^{g-1}~g^{g-1}
\cr}
\eqno(11)$$
where $\delta_2^\prime$ is the radius of the disk containing the isometric
circles $\{I_{T_n},~I_{T_n^{-1}},n=1,...,g\}$.  The conditions for the
fundamental region of the symplectic modular group Sp(2g;Z) lead to a set of
constraints $(Im \tau)_{ss} \ge (Im \tau)_{rr},~s\ge r$, which modify the
integrals over the Schottky group parameters.  The restrictions on the absolute
values of the multipliers that might follow from such constraints are
associated
with a reduction by g!.  As the magnitudes of the multiplier and fixed-point
terms in $(Im \tau)_{nn}$ are of the same order, the conditions on the period
matrix lead to inequalities amongst a larger set of variables.  It is shown in
the appendix that a sequential ordering of $\{(Im~\tau)_{nn}\}$ can be obtained
in a local neighbourhood consisting of J circles, with J bounded, by
restricting
the arguments of $\xi_{1n}-\xi_{2n}$ and $K_n$.  The fixed-point integral will
be reduced by an exponential function of the genus upon repeating the local
ordering throughout the configuration.  A global ordering can be achieved by
choosing a generator associated with each neighbourhood $T_{r_N}$ and imposing
constraints on either $\vert K_{r_N}\vert$ or
$\vert\xi_{1r_N}-\xi_{2r_N}\vert$.
The overall dependence of the combined integral over $\{K_n\}$
and $\{\xi_{1n},~\xi_{2n}\}$ will then be $k_3^g g^{g(1-{1\over J})}$.

The factorial growth is not changed substantially by other terms in the
measure.
Recall that
$$\eqalign{[det(Im~\tau)]^{-13}>&(2\pi)^{13g}\biggl[ln\left({1\over
{\epsilon_0}}\right)
\cr
&~~+{1\over g} \sum_{n=1}^g \sum_\alpha~^{(n,n)} {{\vert\gamma_\alpha\vert^{-2}
\vert\xi_{1n}-\xi_{2n}\vert^2}\over {\vert\xi_{1n}+{{\delta_\alpha}\over
{\gamma_\alpha}}\vert\vert\xi_{2n}+{{\delta_\alpha}\over {\gamma_\alpha}}\vert
\vert\xi_{1n}-V_\alpha\xi_{1n}\vert\vert\xi_{2n}-V_\alpha\xi_{2n}\vert}}
\biggr]^{-13 g}
\cr}
\eqno(12)$$
where $\sum_\alpha~^{(n,n)}$ excludes all elements $V_\alpha$ with left-most or
right-most members equal to $T_n^{\pm 1}$.  In a hexagonal configuration,
corresponding to the densest packing of the isometric circles, the circles
are labelled by the level number $l$, which is related to the distance from the
center of the domain enclosing all $I_{T_n}$ and $I_{T_n^{-1}},n=1,...,g$.
The sum can be separated into a series involving the generators $T_{n_l}$
and a series associated with elements $V_\alpha$ that are products of two or
more generators.  A bound for the first series was obtained in an earlier
investigation [5].

It is shown in the appendix that the upper bound for the series involving
elements $V_\alpha=T_{n_{l_1}}^{\pm 1}...T_{n_{l_m}}^{\pm 1}$ is equal to
${{\delta_0^{\prime 2}}\over {\delta_0^2}}
\left[6 {{\delta_0^{\prime 2}}\over {\delta_0^2}} {{\epsilon_0^\prime}
\over {[1-\epsilon_0^\prime]^2}}\right]^m$, multiplying mth order sums over
fractions consisting of terms with leading-order behaviour
$[l^{T_{n_{l_1}}^{\pm 1}}]^{r_1} ...[l^{T_{n_{l_m}}^{\pm 1}}]^{r_m},~
\sum_{i=1}^m r_i=-(m+2)$, there entire sum is less than that of a convergent
geometric series.  An exponentially decreasing lower bound, therefore, is
sufficient for $[det(Im~\tau)]^{-13}$.

An evaluation of the primitive-element product factors requires an estimate of
$\sum_\alpha~^\prime \vert K_\alpha\vert$ since
$$\eqalign{\prod_\alpha~^\prime &\vert 1-K_\alpha\vert^{-1}~>~exp
\left(-\sum_\alpha~^\prime~\vert K_\alpha\vert\right)
\cr
\prod_\alpha~^\prime&\prod_{p=1}^\infty \vert 1-K_\alpha^p\vert^{-1}~>~
exp\left(-{1\over {1-\epsilon_0^\prime}}\sum_\alpha~^\prime \vert K_\alpha\vert
\right)
\cr}
\eqno(13)$$
A genus-dependent bound can be obtained for the sum
$$\sum_\alpha~^\prime \vert K_\alpha\vert~=~\sum_\alpha~^\prime {{\vert
\gamma_\alpha\vert^{-2}}\over {\left\vert \xi_{1\alpha}+{{\delta_\alpha}\over
{\gamma_\alpha}}\right\vert^2}}~<~2{{\epsilon_0^\prime}\over
{(1-\epsilon_0^\prime)^2}}{{\delta_0^{\prime 2}}\over {\delta_0^2}} g~+~
\sum_{V_{\tilde\alpha}=T_{n_l}V_\beta} {{\vert\gamma_{\tilde\alpha}\vert^{-2}}
\over {\left\vert \xi_{1\tilde\alpha}+{{\delta_{\tilde\alpha}}\over
{\gamma_{\tilde\alpha}}}\right\vert^2}}
\eqno(14)$$
which is demonstrated in the appendix to be only linearly increasing with
respect to the genus as $g\to\infty$.  Therefore, the primitive-element
products are bounded below by an exponentially decreasing function of the
genus, and the lower bound for the integral for configurations of isometric
circles in category (ii) will be reduced further by an exponential factor.

The third category of isometric circles is defined by the parameter q,
$0<q<{1\over 2}$, measuring the fall-off of the distances between the
isometric circles.  The multipliers and the distances between the fixed
points are genus-dependent for these configurations.  The Poincare series
is bounded by
$$\eqalign{\sum_\alpha \vert\gamma_\alpha\vert^{-2}~<~2&\epsilon_0^\prime
\left\vert 1-{{\epsilon_0^\prime}\over {g^{1-2q}}}\right\vert^{-2}
\delta_0^{\prime 2}
\cr
{}~&+~\epsilon_0^{\prime 2}\left\vert 1-{{\epsilon_0^\prime}
\over {g^{1-2q}}}\right\vert^{-4}{{\delta_0^{\prime 4}}\over {g^2}}
\sum_{V_{\tilde\alpha}=T_{n_{l_1}}^{\pm 1} T_{n_{l_2}}^{\pm 1}}
\left\vert {{\delta_{T_{n_{l_1}}^{\pm 1}}}\over {\gamma_{T_{n_{l_1}}^{\pm 1}}}}
+{{\alpha_{T_{n_{l_2}}^{\pm 1}}}\over {\gamma_{T_{n_{l_2}}^{\pm 1}}}}
\right\vert^{-2}
\cr
{}~&+~\sum_{V_{\tilde\alpha}=T_{n_l}V_\beta} \vert
\gamma_{\tilde\alpha}\vert^{-2}
 \cr}
\eqno(15)$$
where $V_\beta$ consists of two elements at least.  An upper bound for the
sum over the elements $V_{\tilde\alpha}=T_{n_{l_1}}^{\pm 1} T_{n_{l_2}}^{\pm
1}$
is found in the appendix to increase linearly with the genus.  This will also
hold true for $\sum_\alpha~^\prime \vert K_\alpha\vert$, and thus, the
primitive-element products will be bounded below by an exponentially decreasing
function of the genus.
\vskip 20pt
The integrals over the multipliers and fixed points depend on q, when
$0<q<{1\over 2}$.  For
 isometric circles defined by the limits in category (iii),
\vskip 15pt
$$\int \prod_{n=1}^g {{d^2K_n}\over {\vert K_n\vert^4}} \vert 1-K_n\vert^4
{}~\ge~\pi^g(1-\epsilon_0^\prime)^{4g}\left({1\over {\epsilon_0^2}}-
{1\over {\epsilon_0^{\prime 2}}}\right)^g g^{2(1-2q)g}$$
\vfill\eject
$$\eqalign{\int {1\over {Vol(SL(2,{\Bbb C}))}}&\prod_{n=1}^g {{d^2\xi_{1n}
d^2\xi_{2n}}\over
{\vert \xi_{1n}-\xi_{2n}\vert^4}}
\cr
&~\ge~{{\vert\xi_{11}^0-\xi_{1g}^0\vert^2\vert\xi_{1g}^0-\xi_{21}^0\vert^2}
\over {\vert\xi_{11}^0-\xi_{21}^0\vert^2}}
\pi^{2g-3}\delta_2^{\prime 2g-4}\left[1-{{(\delta_0+\delta_0^\prime)^2}
\over {4 \delta_2^{\prime 2}g^{1-2q}}}\right]^{g-2}
\cr
&~~~~~\left[{1\over {\delta_0^2}}\left[1+{{\epsilon_0^{1\over 2}}\over
{1+\epsilon_0}}
\left[1-\sum_{n^\prime=1}^\infty {{(2n^\prime-1)!!}\over {(n^\prime +1)!}}
{1\over {2^{n^\prime}}}\right]\right]^{-2} -{1\over {\delta_0^{\prime 2}}}
\right]^{g-1}
g^{2(g-1)q}
\cr}
\eqno(16)$$

There is also a qualitative difference between the lower bound for
$[det(Im~\tau)]^{-13}$ when $q<{1\over 2}$ and the same bound for $q={1\over
2}$.
$$[det(Im~\tau)]^{-13}~>~(2\pi)^{13g}\left[(1-2q)ln~g-ln~\epsilon_0+...
\right]^{-13g}
\eqno(17)$$
Combining these inequalities gives a growth of ${{k^g g^{2(g-gq-q)}}\over
{(ln~g)^{13g}}}$.  Since the multiplier term in $Im~\tau$ grows more rapidly
than
the sum of the logarithms of the fixed-point ratios, the inequalities
${{\epsilon_0}\over {g^{1-2q}}}\le\vert K_g\vert\le \rho^2\vert K_{g-1}\vert,~
...~,\vert K_1\vert\le {{\epsilon_0^\prime}\over {g^{1-2q}}}$ where $\rho$ was
defined
by equation (6).  Division by a factor of $\rho^{-2g}g!$ would lead to an
overall dependence on the genus equal to $(k\cdot e)^g \rho^{2g}
{{g^{(g-2qg-2q)}}\over {(ln~g)^{13g}}}$.  The total contribution of the
configurations in category (iii) can be determined by an overlapping of the
intervals for allowed values of $\vert K_n\vert$,
n=1,...,g, corresponding
to different q.  Selecting a specific value of the genus $g_0$, the upper limit
of one interval can be set equal to the lower limit of the next interval.
Therefore, ${{\epsilon_0}\over {g_0^{1-2q_1}}}={{\epsilon_0^\prime}\over
{g_0}}$, implying that $\epsilon_0=\epsilon_0^\prime g_0^{-2q_1}$, and
${{\epsilon_0}\over {g_0^{1-2q_n}}}={{\epsilon_0^\prime}\over
{g_0^{1-2q_{n-1}}}}$, so that $q_n= nq_1$.  There exists an integer
$N={1\over {2q_1}}+r(q_1),~0\le r(q_1)< 1$ such that $q_N\le {1\over 2}$ and
$q_{N+1}\ge {1\over 2}$.  As $q_1={{ln\left({{\epsilon_0^\prime}\over
{\epsilon_0}}\right)}\over {2ln~{g_0}}}$, for arbitrarily large genus g, the
integer ${{ln~g}\over {ln\left({{\epsilon_0^\prime}\over {\epsilon_0}}\right)}}
+r(\epsilon_0,\epsilon_0^\prime,g)$ equals the number of discrete evenly spaced
values of q in the interval $(0,{1\over 2})$ selected by the requirement of
non-overlapping of the ranges for $\vert K_n\vert$.
The sum of the contributions to the partition function integral associated
with the conditions
${{\epsilon_0}\over {g^{1-2q_i}}}<\vert K_n\vert < {{\epsilon_0^\prime}
\over {g^{1-2q_i}}}$, and ${{\delta_0}\over {g^{q_i}}}< \vert
\xi_{1n}-\xi_{2n}\vert <{{\delta_0^\prime}\over {g^{q_i}}}$, i=1,...,N  is
$${{(k\cdot e)^g \rho^{2g}g^g}\over {(ln~g)^{13g}}}e^{-2
ln({{\epsilon_0^\prime}\over {\epsilon_0}})(g+1)}\left[{{1-g^{-2(g+1)}}\over
{1-e^{-2ln({{\epsilon_0^\prime}\over {\epsilon_0}})(g+1)}}}\right]
\eqno(18)$$
The potential infinity associated with the continuous parameter q
does not occur.

However, each value of q defines a different genus-dependence of the variables
$\vert K_n\vert$ and $\vert\xi_{1n}-\xi_{2n}\vert$.
Magnitudes of the multipliers may be grouped according to the value of q and
inequalities can be imposed within each group.  As the ranges of $\vert
K_n\vert$ do not overlap for different choices of q, the inequalities
$(Im~\tau)_{ss}\ge (Im\tau)_{rr}$ will be satisfied for all $s\ge r$.
Integration of the multipliers in each group can be performed subject to the
constraints on the absolute values, and the integrals will be equal to those
given previously, with a reduced number of variables.   The number of
groupings is the number of ways of
partitioning g ordered objects into
${{ln~g}\over {ln\left({{\epsilon_0^\prime}\over {\epsilon_0}}\right)}}
+r(\epsilon_0,\epsilon_0^\prime,g)$ different sets.  This introduces a
combinatorial factor
$${{\left(g+{{ln~g}\over {ln\left({{\epsilon_0^\prime}\over
{\epsilon_0}}\right)}}-1\right)!}\over {g!\left({{ln~g}\over
{ln\left({{\epsilon_0^\prime}\over {\epsilon_0}}\right)}}-1\right)!}}$$
which can be used to estimate the increase in the lower bound for the
regularized string partition function.

$\underline{3.~A~Separate~Category~of~Isometric~Circles}$

There are other configurations of isometric circles that are consistent with
the genus-independent cut-off imposed on the length of closed geodesics
rendering the moduli space integral finite.  A disk of radius $O(\sqrt{g})$
containing circles of radius O(1) can be projected onto a sphere with handles
of thickness $O({1\over {\sqrt{g}}})$.  As a conformal transformation can be
chosen to change the metric to one of constant negative curvature, so that for
surfaces in the $R=-1$ slice of Teichmuller space, the area increases linearly
with the genus and the thickness of the handles will be expanded by a factor of
$\sqrt{g}$ to O(1) in the intrinsic metric.

Each of the terms in the measure for the integral (2) could be evaluated for
this configuration.  However, $\vert\gamma_\alpha\vert^{-2}$ is of order
O(1) when $V_\alpha=T_{n_1}T_{n_2}$, because the spacing between $I_{T_{n_1}}$
and $I_{T_{n_2}}^{-1}$ is of order O(1) for neighbouring circles.  The Poincare
series $\sum_\alpha \vert\gamma_\alpha\vert^{-2}$ would diverge even at finite
genus.  Bounds for the primitive-element products may exist as the arguments
of the multipliers tend to cancel.  The determinant factor
$[det(Im~\tau)]^{-13}$
also may be bounded below.  It can be seen that the contribution of this
configuration to the partition function is not significant, without requiring
an evaluation of these bounds, because the integrals over the multipliers and
fixed points only increase exponentially as $\vert K_n\vert=O(1)$ and
$\vert\xi_{1n}-\xi_{2n}\vert=O(1)$.

Configurations of isometric circles involving fixed-point distances of
$O(g^r),~r>0$, also do not contribute significantly to the partition function
[5].

$\underline{4.~Upper~Bounds}$

The setting of upper bounds for the integrals over domains in moduli space is
required for a full analysis of the summability of the perturbation series.
Considering the intermediate configurations in category (iii), with
$0<q<{1\over 2}$, one finds
$$\int\prod_{n=1}^g~{{d^2K_n}\over {\vert K_n\vert^4}}\vert 1-K_n\vert^4
{}~\le~\pi^g(1-\epsilon_0)^{4g}\left({1\over {\epsilon_0^2}}-{1\over
{\epsilon_0^{\prime 2}}}\right)^g g^{(1-2q)g}
\eqno(19)$$
$$\eqalign{\int {1\over {Vol(SL(2,{\Bbb
C}))}}\prod_{n=1}^g{{d^2\xi_{1n}d^2\xi_{2n}}\over
{\vert\xi_{1n}-\xi_{2n}\vert^4}}~\le~&{{\vert\xi_{11}^0-\xi_{1g}^0\vert^2
\vert \xi_{1g}^0-\xi_{21}^0\vert^2}\over {\vert\xi_{11}^0-\xi_{21}^0\vert^2}}
\pi^{2g-3}
\cr
&~~\cdot \delta_2^{\prime 2g-4}
\left[{1\over {\delta_0^2}}-{1\over {\delta_0^{\prime 2}}}\right]^{g-1}
g^{2(g-1)q}
\cr}
\eqno(20)$$
The logarithmic increase of the diagonal entries of $Im~\tau$ suggests that
$det(Im~\tau)$ is a rapidly increasing function of the genus.  The determinant
is equal to the volume of the parallelepiped spanned by the basis vectors
${\underline v}_1$,...,${\underline v}_g$ representing the rows of the matrix.
Denoting $\theta_{1...n}$ to be the angle from ${\underline v}_n$ to the
hyperplane spanned by ${\underline v}_1$,...,${\underline v}_{n-1}$ and since
$sin~{\theta_{1...n}}=O(1)$ when the off-diagonal entries are
O(1), it follows that $det(Im~\tau)\ge d(q)^gO(ln~g)^g$ and
$$[det(Im~\tau)]^{-13}~<~(d(q))^{-13g}(2\pi)^{13g}(ln~g+O(1))^{-13g}
\eqno(21)$$
The primitive-element product can be bounded above because
\hfill\break
$\prod_\alpha~^\prime\vert 1-K_\alpha\vert^{-1}
<exp\left(\sum_\alpha~^\prime \vert K_\alpha\vert\right)$, and for each of the
configurations, this function is an exponentially increasing function of the
genus.  It follows that the precise growth of the regularized integral, is
equivalent, up to an exponential factor, for the upper and lower bounds.  The
increase of factorial type with respect to the genus remains unaltered.

$\underline{5.~Graph~Theory~Combinatorics}$

The upper and lower bounds for the partition function have been found by using
the measure defined in terms of multipliers $K_n$ and fixed points $\xi_{1n}$
and $\xi_{2n}$ in the uniformization of Riemann surfaces by Schottky groups.
The Schottky covering is intermediate between the Riemann surface and the
simply connected covering obtained by cutting both the a- and b-cycles, which
are the 2g independent homologically non-trivial paths along the g handles.
Every compact surface of genus $g\ge 2$ can be represented as the quotient
H/G, where H is the upper half plane and G is a discrete subgroup of PSL(2,
${\Bbb R}$)
consisting of hyperbolic elements.  The elements of G are conjugated to
$\left(\matrix{e^{l\over 2}& 0&
         \cr
         0& e^{-{l\over 2}}&
         \cr}
          \right)$, where $l$ can be interpreted as the length of a
corresponding closed geodesic on the Riemann surface [10]. Primitive elements
of the group G, which are not powers of the other elements, correspond to
simple
closed geodesics.  Denoting the primitive elements by the index set
$\{\gamma\}$, the length spectrum $\{l_\gamma\}$ always has a minimum value
$l_0$ for compact surfaces.  The Selberg trace function is defined by
$$Z(s)~=~\prod_\gamma~^\prime \prod_{n=0}^\infty~[1-e^{-(s+n)l_\gamma}]
\eqno(22)$$
The partition function is $\sum_{g=0}^\infty \kappa^g \int_{M_g} d\mu_{WP}
Z(2)Z^\prime (1)^{-13}$ where $M_g$ is the moduli space of genus g surfaces and
$d\mu_{WP}$ is the Weil-Petersson measure.  If the minimum length $l_0$ is
bounded below by
 ${\bar l_0}$
in
${\bar M_g}-N(D_g)$
where ${\bar M}_g$ is the closure of $M_g$
and $N(D_g)$ is a neighbourhood of the compactification  divisor, it
can be shown that
\hfil\break
 $Z(2)Z^\prime (1)^{-13} >c_1({\bar l_0})c_2({\bar l_0})^g$.  The
integral over the Weil-Petersson measure in a region in Teichmuller space can
be estimated.  Any Riemann surface of finite genus can be viewed as a thickened
trivalent graph with branches at the final level intertwined.  The number of
different pairings of the 2g branches is (2g-1)!!.  This is consistent with the
counting of non-isomorphic trivalent graphs [11].

The number of cells may increase at a factorial rate, but it remains to be
shown that
the cells are non-intersecting and that they belong to a single fundamental
region of the symplectic group.  This requires a study of the transformation
between surfaces formed out of graphs with different branches intertwined.  If
the transformation does not lie in the identity component of the diffeomorphism
group, then the surfaces lie in different fundamental domains of the mapping
class group, and both cannot be included in the counting.  Integration
over the length and twist parameters suggests that the volume occupied by each
cell depends exponentially on the genus, so that the growth will be determined
by the restriction to the fundamental domain of the modular group.

In this connection, it may be observed that a trivalent graph in a plane will
have intersecting branches at large order if the length of the branches is
fixed at a constant value.  If the graph is then thickened to obtain a surface
in ${\Bbb R}^3$ with minimal genus-independent cross-sectional area, the disks
at the ends of
the branches at the final level will lie in overlapping cylindrical regions,
and it will not be possible to join the pairs of disks without joining other
disks.  This property, which would continue to hold in ${\Bbb R}^{26}$,
suggests that the counting of closed surfaces at finite genus
would changed by the overlapping of the regions.
 Intersection of the branches may be avoided if
the lengths of the branches are allowed to increase exponentially  with respect
to the order.   An exponential increase would reduce the integration over the
length parameters of all of the g handles by a factor which would affect the
factorial growth obtained by
counting of the intertwined graphs.  An exponential decrease of the length of
the branches would also eliminate intersections, but it is not consistent with
a minimal genus-independent cross-sectional area.

Thus, the factorial growth derived from graph theory appears to be linked to
the
emergence of a point-like structure associated with the boundaries of the
string
worldsheet in the infinite-genus limit.  If the constraint
$\vert\gamma_n\vert^{-2} \sim {1\over g}$ is not imposed on the Schottky group
parameters, then
a factorial increase can be easily obtained through integration of the
variables $\vert K_n\vert$ or $\vert \xi_{1n}-\xi_{2n}\vert$ [6].  The surfaces
giving rise to these divergences in this limit, having boundaries of zero
linear
measure, are eliminated by the
genus-independent cut-off on the lengths of the closed geodesics.  The
reduction
in the growth of the string integral has been verified in the calculations
of the bounds in Sections 2 and 4, using the Schottky group parametrization.

$\underline{6.~The~Partition~Function~for~a~Curved~Background}$

The partition function for a string propagating in a curved background would be
$$Z~=~\sum_{g=0}^\infty \kappa^g \int {{D[h_{\alpha\beta}]D[X^\mu]}\over
{Vol(Diff~\Sigma_g)Vol(Conf~\Sigma_g)}}e^{-{1\over 2}\int d^2\xi \sqrt{h}
h^{\alpha\beta}\partial_\alpha X^\mu \partial_\beta X^\nu g_{\mu\nu}}
\eqno(23)$$
Formulas for partition functions of closed bosonic strings have been obtained
for conformally flat backgrounds [12] and group manifolds [13].

Divergences in the partition function may be identified with particular factors
in the measure.  In flat space, the general bosonic N-string g-loop amplitude
can be obtained by sewing together g pairs of legs of an (N+2g)-string tree
amplitude.  The integrand of the N-string amplitude can be written as
$$d^{N-3}W_N(X,b,c)~=~\int d^N\hat{c} e^{-S(X,b,c+\hat{c})}(2 \pi)^{26}
\delta(\sum_{i=1}^N p_i)
\eqno(24)$$
where $p_i$ are the string momenta, S(X,b,c) is the string action involving
coordinate and ghost fields, and the ghost modes $\hat c$ are needed to
reproduce Koba-Nielsen-type variables.  An M-string vertex can be sewed with
an N-string vertex along one of the legs to give an (M+N-2)-string-tree
amplitude with integrand
$$\eqalign{d^{(M+N-5)}W_{M+N-2}(X+TX^\prime &,b+Tb^\prime,c+Tc^\prime)
\cr
{}~=~(2\pi)^{26}\int D[X_s,& b_s,c_s] dc_s \delta^{26}(p_s) e^{-S[x_s,b_s,c_s+
\hat{c_s}]}
\cr
d^{(M-3)}& W_M(X,b,c)d^{(N-3)}W_N(X^\prime,b^\prime,c^\prime)
\cr}
\eqno(25)$$
with $X_s,~b_s,~c_s$ and $\hat{c_s}$ being sewing fields and T being the
projective transformation connecting the two legs [8].  Each sewing of a pair
of legs will introduce one modular integration variable through the sewing
field $c_s$.  Sewing g pairs of legs of a 2g-string tree amplitude will involve
3g-3 integration variables for moduli space.  The projective transformations
for
g sewings generate the Schottky group uniformizing the Riemann surface.
Several of the factors in the measure can be traced to integration of the
sewing fields.  The string sewing fields $\{X_{sn}\}$ give the factors
$(det~Im~\tau)^{-13} \prod_\alpha~^\prime \prod_{p=1}^\infty
\vert 1-K_\alpha^p\vert^{-52}$, while the ghost sewing fields
$\{b_{sn},c_{sn}\}$ give $\prod_\alpha~^\prime \prod_{p=1}^\infty \vert
1-K_\alpha\vert^4$.  The remaining factors may be obtained through integration
over the ghost zero modes.

It has been shown that divergences in the regularized partition function result
from factors associated with the ghost zero modes, although the logarithmic
factor, which could affect summability of the perturbation series, arises from
$(det~Im~\tau)^{-13}$.  In curved space, a similar derivation of the factor in
the measure might be feasible, with the path integration being modified by the
change in the metric.

Without explicitly deriving the measure, it can be seen that the finite volume
of a compact target space would lead to a cut-off in the volume of the string
world-sheet if the mapping  between the target-space metric and the intrinsic
world-sheet metric is differentiable and invertible.  This would result in the
exclusion of certain configurations of isometric circles in the large-genus
limit.  It has been demonstrated here that the classification of these
configurations is related to the type of divergence in the perturbation series.
Thus, the inclusion of only particular configurations of isometric circles
for a world-sheet embedded in a compact curved target manifold would alter the
Borel summability of the series.
\vskip 5pt
\centerline{\it Acknowledgements}
This research was initiated at the I.C.T.P., Trieste, where financial support
was received from the Atomic Energy Commission, Vienna.
\vfill
\eject

\centerline{\bf Appendix}

For the second category of isometric circles, $\vert\xi_{1n}-\xi_{2n}\vert
=O({1\over {\sqrt{g}}})$, so that the dominant contribution to the fixed-point
sum in $Im~\tau$ arises from the group of neighbouring isometric circles
$I_{V_\alpha}$ such that $\left\vert\xi_{1s}+{{\delta_\alpha}\over
{\gamma_\alpha}}\right\vert=O({1\over{\sqrt{g}}})$ and $\vert \xi_{1s}-V_\alpha
\xi_{1s}\vert=O({1\over {\sqrt{g}}})$.  The fixed-point term in $Im~\tau$
will then be O(1), which is the same order of magnitude as
$ln~\left({1\over {\vert K_n\vert}}\right)$.  Thus, inequalities amongst the
fixed-point variables and even the arguments of the multipliers may be used
as alternative representations of the conditions $(Im~\tau)_{ss}\ge
(Im~\tau)_{rr},~s\ge r$.

Moreover, the radii $\vert\gamma_\alpha\vert^{-1}$ decrease as the number of
generators in $V_\alpha$ increases, and an even distribution of the phases of
$\gamma_\alpha^{-2}$ in the interval $[-\pi,\pi]$ leads to a cancellation of
terms in the sum.  It will be sufficient, therefore, to consider the
isometric circles $I_{T_{n_l}^{(j)}}$ of the fundamental generators, where
l is the level number in the hexagonal configuration and (j) labels the
6l-6 circles at each level.
Let
$$\eqalign{C_{l,s}^{(j)}~=&~{{\vert \xi_{1s}-\xi_{2s}\vert^2}\over
{\left\vert\xi_{1s}
+{{\delta_{T_{n_l}^{(j)}}}\over {\gamma_{T_{n_l}^{(j)}}}}\right\vert
\left\vert \xi_{2s}+{{\delta_{T_{n_l}^{(j)}}}\over
{\gamma_{T_{n_l}^{(j)}}}}\right\vert
\vert\xi_{1s}-T_{n_l}^{(j)} \xi_{1s}\vert
\vert\xi_{2s}-T_{n_l}^{(j)}\xi_{2s}\vert}}
\cr
D_{l,s}^{(j)}~=&~\vert\gamma_{n_l}^{(j)}\vert^{-2}C_{l,s}^{(j)}
cos\left[arg\left({{K_{n_l}^{(j)} (\xi_{2s}-T_{n_l}^{(j)}\xi_{2s})^{-1}}
\over {(1-K_{n_l}^{(j)})^2 \left(\xi_{1s}+{{\delta_{n_l}^{(j)} }\over
{\gamma_{n_l}^{(j)}}}\right)\left(\xi_{2s}+{{\delta_{n_l}^{(j)}}\over
{\gamma_{n_l}^{(j)}}}\right)(\xi_{1s}-T_{n_l}^{(j)}\xi_{1s})}}\right)
\right]
\cr
E_{l,s}^{(j)}~=&~\sqrt{\vert\gamma_{n_l}^{(j)}\vert^{-4}C_{l,s}^{(j)2}
-D_{l,s}^{(j)2}}
\cr}
\eqno(A.1)$$
Then, if two pairs of fixed points $(\xi_{1r},\xi_{2r})$ and
$(\xi_{1s},\xi_{2s})$ are located near to each other, and if the difference in
the arguments of $(\xi_{1r}-\xi_{2r})$ and $(\xi_{1s}-\xi_{2s})$ is denoted
by $\epsilon_{rs}$, the inequalities $(Im~\tau)_{ss}\ge(Im~\tau)_{rr},~s\ge r$
will be satisfied when
$$\eqalign{\epsilon_{rs}\ge {1\over 2}arc~sin&\biggl[{1\over
2}\biggl[\sum_{l=2}^J
\sum_{{j=1}\atop {[l(j)]\not=r,s]}}^{6l-6}
[D_{l,s}^{(j)}sin[2 arg(\xi_{1l}^{(j)}-\xi_{2l}^{(j)})+2arg(\xi_{1s}-\xi_{2s})]
\cr
&+E_{l,s}^{(j)}cos[2
arg(\xi_{1l}^{(j)}-\xi_{2l}^{(j)})+2arg(\xi_{1s}-\xi_{2s})]]
\biggr]^{-1} \cdot ln\left({{\epsilon_0^\prime}\over
{\epsilon_0}}\right)\biggr]
\cr}
\eqno(A.2)$$
where J is bounded, because the dominant contribution to the fixed point sum
corresponds to the isometric circles $I_{T_{n_l}^{(j)}}$ and
$I_{T_{n_l}^{(j)}}^{-1}$ near to the circles $I_{T_s}$ and $I_{{T_s}^{-1}}$.
Since $\vert \gamma_{n_l}^{(j)}\vert^{-2}=O({1\over g})$, $C_{l,s}^{(j)}=O(g),~
D_{(l,s)}^{(j)}=O(1)$ and $E_{l,s}^{(j)}=O(1)$, $\epsilon_{rs}=O(1)$.

It may be assumed that the range of arguments and absolute values of
$K_{n_l}^{(j)}$ can be selected to be sufficiently narrow so that it will not
affect the inequalities for the fixed-point ratios significantly.  Given a
range of values for $arg(\xi_{1s}-\xi_{2s})$ and a choice for
$arg(\xi_{1l}^{(j)}-\xi_{2l}^{(j)})$, the shifted range of values for
$arg(\xi_{1r}-\xi_{2r})$ can be obtained.  Therefore, the ranges of the
arguments $arg(\xi_{1r_n}-\xi_{2r_n})$ can be ordered sequentially.  An
alternative method, which allows $arg(\xi_{1l}^{(j)}-\xi_{2l}^{(j)})$ to be
arbitrary initially, would involve a coupled set of equations resulting from
inequalities of the type given in equation (A.2).  As a consequence of the
width of the angular intervals being of order O(1), the fixed point integral
is reduced by an exponential factor.

The bound for the determinant factor $[det(Im~\tau)]^{-13}$ has been given
in Section 2.  The fixed-point sum in the lower bound can be shown to be
finite for arbitrary genus [5].  The sum can be expressed as
$$\eqalign{\sum_{n_l\not= n}&
{{\vert\gamma_{n_l}\vert^{-2}\vert\xi_{1n}-\xi_{2n}\vert^2}
\over {\vert\xi_{1n}-T_{n_l}\xi_{1n}\vert\vert\xi_{2n}-T_{n_l}\xi_{2n}\vert
\left\vert\xi_{1n}+{{\delta_{n_l}}\over {\gamma_{n_l}}}\right\vert
\left\vert\xi_{2n}+{{\delta_{n_l}}\over {\gamma_{n_l}}}\right\vert}}
\cr
&+\sum_{n_l\not=n}{{\vert\gamma_{n_l}\vert^{-2}\vert\xi_{1n}-\xi_{2n}\vert^2}
\over
{\vert\xi_{1n}-T_{n_l}^{-1}\xi_{1n}\vert\vert\xi_{2n}-T_{n_l}^{-1}\xi_{2n}
\vert \left\vert\xi_{1n}-{{\alpha_{n_l}}\over {\gamma_{n_l}}}\right\vert
\left\vert\xi_{2n}-{{\alpha_{n_l}}\over {\gamma_{n_l}}}\right\vert}}
\cr
&+\sum_{V_{\tilde\alpha}=T_{n_l}V_\beta}~^{(n,n)}
{{\vert\gamma_{\tilde\alpha}\vert^{-2}\vert\xi_{1n}-\xi_{2n}\vert^2}\over
{\vert\xi_{1n}-V_{\tilde\alpha}\xi_{1n}\vert\vert\xi_{2n}-V_{\tilde\alpha}
\xi_{2n}\vert \left\vert\xi_{1n}+{{\delta_{\tilde\alpha}}\over
{\gamma_{\tilde\alpha}}}\right\vert
\left\vert\xi_{2n}+{{\delta_{\tilde\alpha}}\over
{\gamma_{\tilde\alpha}}}\right\vert}}
\cr}
\eqno(A.3)$$
Suppose that $\xi_{1n}$ and $\xi_{2n}$ lie at the center of a hexagonal
arrangement of 2g isometric circles spaced apart from each other by a distance
of $O({1\over {\sqrt{g}}})$.  Since
$$\eqalign{&\left\vert\xi_{2n}+{{\delta_{n_l}}\over {\gamma_{n_l}}}\right\vert
{}~\ge~(l-1){{\delta_0}\over
{\sqrt{g}}}~~~~~\left\vert\xi_{1n}+{{\delta_{n_l}}\over
{\gamma_{n_l}}}\right\vert~\ge~ max\left\{\left[l-1-{{\delta_0^\prime}\over
{\delta_0}}\right]{{\delta_0}\over {\sqrt{g}}}, {{\delta_0}\over
{\sqrt{g}}}\right\}
\cr
&\vert\xi_{2n}-T_{n_l}\xi_{2n}\vert~\ge~max
\left\{\left[l-1-{{\delta_0^\prime}\over
{\delta_0}}\left[1+{{2\epsilon_0^{\prime{1\over 2}}}\over
{1-\epsilon_0^\prime}}
\right]\right]{{\delta_0}\over {\sqrt{g}}}, {{\delta_0}\over {\sqrt g}}\right\}
\cr
&\vert\xi_{1n}-T_{n_l}\xi_{1n}\vert~\ge~max\left\{\left[l-1-2{{\delta_0^\prime}\over {\delta_0}}\left[1+{{\epsilon_0^{\prime{1\over 2}}}\over {1-\epsilon_0^\prime}}
\right]\right]{{\delta_0}\over {\sqrt{g}}},{{\delta_0}\over {\sqrt g}}\right\}
\cr}
\eqno(A.4)$$
and as there are only 6(l-1) circles at level l in a configuration with
$[{1\over 2}+{1\over 6}\sqrt{9+24g}]$ levels, the first two sums in (A.3)
converge as $g\to\infty$.

For the sum over elements $V_{\tilde\alpha}=T_{n_l}V_\beta$, the ratio of the
square of the radii of $I_{V_{\tilde\alpha}}$ and $I_{V_\beta}$ is
$$\left\vert{{\gamma_{\tilde\alpha}}\over {\gamma_\beta}}\right\vert^{-2}
{}~=~\vert\gamma_{n_l}\vert^{-2}\left\vert{{\delta_{n_l}}\over {\gamma_{n_l}}}
+{{\alpha_\beta}\over {\gamma_\beta}}\right\vert^{-2}~<~
{{\epsilon_0^\prime}\over {[1-\epsilon_0^\prime]^2}}{{\delta_0^{\prime 2}}\over
g}
\left\vert{{\delta_{n_l}}\over {\gamma_{n_l}}}+{{\alpha_\beta}\over
{\gamma_\beta}}\right\vert^{-2}
\eqno(A.5)$$
It follows that the coefficients multiplying the convergent jth order sums of
fractions involving the level numbers are ${{\delta_0^{\prime 2}}\over
{\delta_0^2}}\left[6
{{\delta_0^{\prime 2}}\over {\delta_0^2}}{{\epsilon_0^\prime}\over
{(1-\epsilon_0^\prime)^2}}\right]^j$,
and the entire fixed point sum containing $\xi_{1n}$ and $\xi_{2n}$ is bounded
by a convergent geometric series if $6{{\delta_0^{\prime 2}}\over {\delta_0^2}}
{{\epsilon_0^\prime}\over {[1-\epsilon_0^\prime]^2}}<\Delta$, where
$\Delta$ is a constant determined by the level number sums.

A similar method can be used for estimating the sum $\sum_\alpha~^\prime \vert
K_\alpha\vert$.  The first set of elements included in the sum over
$V_{\tilde\alpha}=T_{n_l}V_\beta$ in equation (14) is of the form
$\{T_{n_{l_1}}
T_{n_{l_2}}\}$.  For these elements $\xi_{1\tilde\alpha}\in
D_{V_{\tilde\alpha}^{-1}} \subset D_{T_{n_{l_1}}}^{-1}$ and
$-{{\delta_{\tilde\alpha}}\over {\gamma_{\tilde\alpha}}}\in
D_{V_{\tilde\alpha}}
\subset D_{T_{n_{l_2}}}$.  Denoting the level of $I_{T_{n_{l_1}}}$ to be $l$
and the level of $I_{T_{n_{l_2}}^{-1}}$ to be $l+l_0$, it follows that
$\left\vert\xi_{1\tilde\alpha}+{{\delta_{\tilde\alpha}}\over
{\gamma_{\tilde\alpha}}}\right\vert\ge {{\delta_0}\over {\sqrt{g}}}$ when
\vskip 1pt
\noindent
$\vert l_0\vert\le \left\{ 2{{\delta_0^\prime}\over
{\delta_0}}\left(1+{{2\epsilon_0^{\prime{1\over 2}}}\over
{1-\epsilon_0^\prime}}\right) \right\}$ and $\left\vert\xi_{1\tilde\alpha}
+{{\delta_{\tilde\alpha}}\over {\gamma_{\tilde\alpha}}}\right\vert
\ge \left[\vert l_0\vert-2{{\delta_0^\prime}\over {\delta_0}}\left(1+
{{2\epsilon_0^{\prime {1\over 2}}}\over {1-\epsilon_0^\prime}}\right)\right]
{{\delta_0}\over {\sqrt{g}}}$ when
\vskip 1pt
\noindent
$\vert l_0\vert\ge \left\{2{{\delta_0^\prime}\over
{\delta_0}}\left(1+{{2\epsilon_0^{\prime {1\over 2}}}\over
{1-\epsilon_0^\prime}}
\right)\right\}+1$.
Then
\vskip 3pt
$$\vert\gamma_{\tilde\alpha}\vert^{-2}~=~\vert\gamma_{n_{l_1}}\vert^{-2}
\vert\gamma_{n_{l_2}}\vert^{-2}\left\vert{{\delta_{n_{l_1}}}\over
{\gamma_{n_{l_1}}}}+{{\alpha_{n_{l_2}}}\over
{\gamma_{n_{l_2}}}}\right\vert^{-2}
\le {{\epsilon_0^{\prime 2}}\over {[1-\epsilon_0^\prime]^4}}
{{\delta_0^{\prime 4}}\over {g^2}}\left\vert {{\delta_{n_{l_1}}}\over
{\gamma_{n_{l_1}}}}+{{\alpha_{n_{l_2}}}\over
{\gamma_{n_{l_2}}}}\right\vert^{-2}
\eqno(A.6)$$
\vskip 3pt
\noindent
Let $d_l$ be the distance from an isometric circle at level $l$ to the center
of the configuration.  If the isometric circles at level $l+l_0$ are labelled
by the index j, a sum of inverse squares of distances from $I_{T_{n_{l_1}}}$
to $I_{T_{n_{l+l_0}}^{(j)}}$ is required by equations (14) and (A.6).
\vfill\eject
\noindent
When $l+l_0 \gg 1$, this sum may be approximated by the integral
$$2\int{{dx}\over {\left[d_{l+l_0}^2+d_l^2-2d_ld_{l+l_0} cos \left({{x\pi}
\over {3(l+l_0-1)}}\right)\right]}}~=~{{6(l+l_0-1)}\over {d_{l+l_0}^2-d_l^2}}
\eqno(A.7)$$
so that the following bound
$$\eqalign{18{{\epsilon_0^{\prime 2}}\over {[1-\epsilon_0^\prime]^4}}
{{\delta_0^{\prime 4}}\over {\delta_0^4}}
\sum_{l=2}^{[{1\over 2}+{1\over 6}\sqrt{9+24g}]} [l-1]&
\left[1+{1\over {2l}}+\sum_{{l_0=-l+2}\atop {l_0\not=0}}
^{\left\{2{{\delta_0^\prime}\over {\delta_0}}
\left[1+{{2\epsilon_0^{\prime{1\over 2}}}\over {1-\epsilon_0^\prime}}\right]
\right\}}
\left\vert{1\over {l_0}}+{1\over {2l+l_0}}\right\vert\right]
\cr
+18{{\epsilon_0^{\prime 2}}\over {[1-\epsilon_0^\prime]^4}}
{{\delta_0^{\prime 4}}\over {\delta_0^4}}\sum_{l=2}^{[{1\over 2}+{1\over 6}
\sqrt{9+24g}]} [l-1]&
\cr
\sum_{l_0=\left\{2{{\delta_0^\prime}\over
{\delta_0}}\left[1+{{2\epsilon_0^{\prime
{1\over 2}}}\over {1-\epsilon_0^\prime}}\right]\right\}+1}^{[-{1\over
2}+{1\over 6}
\sqrt{9+24g}]-1}&
\left[l_0-2{{\delta_0^\prime}\over {\delta_0}}\left[1+{{2\epsilon_0^{\prime
{1\over 2}}}\over {1-\epsilon_0^\prime}}\right]\right]^{-2}
\left[{1\over {l_0}}+{1\over {2l+l_0}}\right]
\cr}
\eqno(A.8)$$
is obtained as $d_l\ge [l-1]{{\delta_0}\over {\sqrt{g}}}$.  For large g, the
sum grows as O(g), and since the higher-order terms may be bounded similarly,
 so that the entire sum $\sum_\alpha~^\prime \vert K_\alpha\vert$
is bounded above by a function that increases only linearly with the genus as
$g\to\infty$.

For the third category of isometric circles, the estimates can be found for
each of the sums in equation (15).  The upper bound for $\sum_\alpha~^\prime
\vert K_\alpha\vert$ would be affected by a change in the range of distances
$\vert\xi_{1n}-\xi_{2n}\vert,~n=1,...,g$.  When
$V_{\tilde\alpha}=T_{n_{l_1}}^{\pm 1}T_{n_{l_2}}^{\pm 1}$,
the distances $\left\vert
\xi_{1\tilde\alpha}+{{\delta_{\tilde\alpha}}\over
{\gamma_{\tilde\alpha}}}\right\vert$ are not correlated as closely with the
distances $\left\vert{{\delta_{T_{n_{l_1}}^{\pm 1}}}\over
{\gamma_{T_{n_{l_1}}^{\pm1}}}}+{{\alpha_{T_{n_{l_2}}^{\pm 1}}}\over
{\gamma_{T_{n_{l_2}}^{\pm 1}}}}\right\vert$.  Suppose that $j_1$ and $j_2$
index
the isometric circles $\{I_{T_{n_{l_1,j_1}}}, j_1=0,1,...,6(l-1)-1\}$ at level
$l$ and the circles $\{I_{T_{n_{l_2,j_2}}^{-1}}, j_2=0,1,..., 6(l+l_0-1)-1\}$
at
level $l+l_0$.  When $V_{\tilde\alpha}=T_{n_{l_1}}T_{n_{l_2}}$,
$$\left\vert{{\delta_{T_{n_{l_1}}^{(j_1)}}}\over
{\gamma_{T_{n_{l_1}}^{(j_1)}}}}
+{{\alpha_{T_{n_{l_2}}^{(j_2)}}}\over
{\gamma_{T_{n_{l_2}}^{(j_2)}}}}\right\vert^{-2}
\le \left[d_{l+l_0}^2+d_l^2-2d_ld_{l+l_0} cos\left({{i(j_1,j_2)\pi}\over
{3(l+l_0-1)}}\right)\right]^{-1}
\eqno(A.9)$$
where $i(j_1,j_2)=0,1,..., 6(l+l_0-1)-1$, but $d(I_{T_{n_{l_1,j_1}}^{-1}},~
I_{T_{n_{l_2, j_2}}})$ can range from ${{\delta_0}\over {\sqrt{g}}}$ to
$2{{\delta_0^\prime}\over {g^q}}+\left\vert{{\delta_{T_{n_{l_1}}^{(j_1)}}}
\over {\gamma_{T_{n_{l_1}}^{(j_1)}}}}+{{\alpha_{T_{n_{l_2}}^{(j_2)}}}\over
{\gamma_{T_{n_{l_2}}^{(j_2)}}}}\right\vert$.  When $\vert l_0\vert$ lies in the
range
$[0, 2{{\delta_0^\prime}\over {\delta_0}}g^{{1\over 2}-q}]$, $\left\vert
\xi_{1\alpha}+{{\delta_\alpha}\over {\gamma_\alpha}}\right\vert$ can assume its
minimum value ${{\delta_0}\over {\sqrt{g}}}$.  Beyond this range
$$\left\vert\xi_{1\alpha}+{{\delta_\alpha}\over {\gamma_\alpha}}\right\vert
\ge \vert l_0\vert {{\delta_0}\over {\sqrt{g}}}-2{{\delta_0^\prime}\over {g^q}}
\left[1+{{2\epsilon_0^{\prime {1\over 2}}}\over {\left(1-{{\epsilon_0^\prime}
\over {g^{1-2q}}}\right)}}\right]
\eqno(A.10)$$
Replacing the limit of the first sum over $l_0$ in equation (A.8) by
$\left\{ 2{{\delta_0^\prime}\over {\delta_0}}g^{{1\over 2}-q} \right\}$ leads
to a bound of
the form $6{{\epsilon_0^{\prime 2}}\over
{(1-\epsilon_0^\prime)^4}}{{\delta_0^{\prime 4}}\over {\delta_0^4}}({1\over
2}-q)
g~ln~g$ for $\sum_\alpha~^\prime \vert K_\alpha\vert$.  This modification
would be sufficient to obtain an upper bound for the primitive-element products
that increases at a factorial rate with respect to the genus, and it is
necessary to refine the estimate of the sum.
 By considering the
densest packing of $6(l+l_0-1)$ circles $\{I_{T_{n_{l_2}}}\}$ about
$I_{T_{n_{l_1,j_1}}^{-1}}$, one finds that the distances
$[d(I_{T_{n_{l_1,j_1}}^{-1}}, \{I_{T_{n_{l_2,j_2}}}\})]^{-2}$ range from
${{\delta_0}\over {\sqrt{g}}}$ to $[-{1\over 2}+{1\over 2}\sqrt{1+8(l+l_0-1)}]
{{\delta_0}\over {\sqrt{g}}}$. Using the average value of
$[d(I_{T_{n_{l_1,j_1}}^{-1}}, \{I_{T_{n_{l_2,j_2}}}\})]^{-2}$, which is
approximately ${g\over {2\delta_0^2}}{{ln(l+l_0-1)}\over {l+l_0-1}}$, it
follows
that the sum over elements
$V_{\tilde\alpha}=T_{n_{l_1}}^{\pm 1}T_{n_{l_2}}^{\pm 1}$ in the upper bound
for $\sum_\alpha~^\prime \vert K_\alpha \vert$
will be less than
$$\eqalign{\epsilon_0^{\prime 2}&\left[1-{{\epsilon_0^\prime}\over
{g^{1-2q}}}\right]^{-4}{{\delta_0^{\prime 4}}\over {\delta_0^4}}
\cr
&\biggl[\sum_{l=2}^{[{1\over 2}+{1\over 6}\sqrt{9+24g}]} 18[l-1]
\left[1+{1\over {2l}}+{1\over 2}\sum_{{l_0=-l+2}\atop {l_0 \not= 0}}^{\left\{
2{{\delta_0^\prime}\over {\delta_0}}g^{{1\over 2}-q}\right\} }
{{ln(l+l_0-1)}\over {l+l_0-1}}
\left\vert{1\over l_0}+{1\over {2l+l_0}}\right\vert\right]
\cr
&+\sum_{l=2}^{[{1\over 2}+{1\over 6}\sqrt{9+24g}]} 18[l-1]
\sum_{l_0=\left\{ 2{{\delta_0^\prime}\over {\delta_0}}g^{{1\over 2}-q}\right\}
+1}^{[-{1\over 2}
+{1\over 6}\sqrt{9+24g}]-1} \left[l_0-2{{\delta_0^\prime}\over {\delta_0}}
g^{{1\over 2}-q}\right]^{-2}\left[{1\over {l_0}}+{1\over
{2l+l_0}}\right]\biggr]
\cr}
\eqno(A.11)$$
 Evaluation of the factor
$\left\vert\xi_{1\tilde\alpha}+{{\delta_{\tilde\alpha}}
\over {\gamma_{\tilde\alpha}}}\right\vert$ implies that the contribution
of the generators $\{I_{T_{n_l}^{\pm1}}\}$
 to $\sum_\alpha~^\prime \vert K_\alpha\vert$  will be of order $O(g^{2q})$,
indicating, together with equation (A.11) that the sum will be bounded by a
linearly increasing function of the genus.
The primitive-element product factors should not alter significantly the growth
of the bound obtained after integration over the multipliers and fixed-points.
\vfill
\eject
\centerline{\bf References}
\item{[1]}  D. Friedan and S. Shenker, Phys. Lett. ${\underline {B175}}$ (1986)
287
\hfil\break
D. Friedan and S. Shenker, Nucl. Phys. ${\underline {B281}}$ (1987) 509
\vskip 3pt
\item{[2]}  D. J. Gross `Non-Perturbative String Theory' in ${\underline
{Random~ Surfaces~and~Quantum}}$
\hfil\break
${\underline {Gravity}}$,
ed. by O. Alvarez et. al., (Plenum Press, New York, 1991) 255-267
\vskip 3pt
\item{[3]} D. J. Gross and V. Periwal, Phys. Rev. Lett. ${\underline {60}}$
(1988) 2105
\vskip 3pt
\item{[4]} S. Mandelstam, `The Interacting String Picture and Functional
Integration'
\hfil\break
Proceedings of the Workshop on Unified String Theories, ITP-Santa Barbara, ed.
by
\hfil\break
 M. Green and D. Gross (Singapore: World Scientific, 1986) 46-102
\vskip 3pt
\item{[5]}  S. Davis, ICTP preprint (1992) IC/92/431
\vskip 3pt
\item{[6]}  S. Davis, University of Cambridge preprint (1994) DAMTP-R/94/1
\vskip 3pt
\item{[7]}  C. L. Siegel, ${\underline {Topics~in~Complex~Function~Theory}}$,
Vol.3
\hfil\break
(New York: Wiley, 1973)
\vskip 3pt
\item{[8]}  J. L. Petersen, K. O. Roland and J. R. Sidenius, Phys. Lett.
 ${\underline {B205}}$ (1988) 262-266
\hfil\break
J. L. Petersen and J. R. Sidenius, Nucl. Phys. ${\underline {B301}}$ (1988)
 247-266
\hfil\break
K. O. Roland, Nucl. Phys. ${\underline {B313}}$ (1989) 432-446
\vskip 3pt
\item{[9]}  S. Davis, Class. Quantum Gravity ${\underline 7}$ (1990) 1887-1893
\vskip 3pt
\item{[10]}  E. D'Hoker and D. H. Phong, Rev. Mod. Phys. Vol. ${\underline
{60}}$, No. 4
(1988) 917-1065
\vskip 3pt
\item{[11]}  B. Bollobas, J. Lond. Math. ${\underline {26}}$ (1982) 201-206
\vskip 3pt
\item{[12]}  A. A. Tseytlin, Int. J. Mod. Phys. A Vol. ${\underline {5}}$, No.
4 (1990)
589-658
\vskip 3pt
\item{[13]}  D. Gepner and E. Witten, Nucl. Phys. ${\underline {B278}}$ (1986)
493-549
\vfill
\eject
\end